\documentclass[conference]{IEEEtran}
\IEEEoverridecommandlockouts

\usepackage{cite}
\usepackage{amsmath,amssymb,amsfonts}
\usepackage{algorithmic}
\usepackage{graphicx}
\usepackage{lipsum}  
\usepackage{amsthm}
\usepackage{enumitem,kantlipsum}
\usepackage{url}

\begin{document}

\title{Federated Learning Approach to Mitigate Water Wastage}

\author{\IEEEauthorblockN{Sina Hajer Ahmadi}
\IEEEauthorblockA{\textit{Department of Electrical and Computer Engineering} \\
\textit{University of Waterloo}}
\and
\IEEEauthorblockN{Amruta Pranadika Mahashabde}
\IEEEauthorblockA{\textit{Department of Electrical and Computer Engineering} \\
\textit{University of Waterloo}}
}

\maketitle

\begin{abstract}
Residential outdoor water use in North America accounts for nearly 9 billion gallons daily, with approximately 50\% of this water wasted due to over-watering, particularly in lawns and gardens. This inefficiency highlights the need for smart, data-driven irrigation systems. Traditional approaches to reducing water wastage have focused on centralized data collection and processing, but such methods can raise privacy concerns and may not account for the diverse environmental conditions across different regions. In this paper, we propose a federated learning-based approach to optimize water usage in residential and agricultural settings. By integrating moisture sensors and actuators with a distributed network of edge devices, our system allows each user to locally train a model on their specific environmental data while sharing only model updates with a central server. This preserves user privacy and enables the creation of a global model that can adapt to varying conditions. Our implementation leverages low-cost hardware, including an Arduino Uno microcontroller and soil moisture sensors, to demonstrate how federated learning can be applied to reduce water wastage while maintaining efficient crop production. The proposed system not only addresses the need for water conservation but also provides a scalable, privacy-preserving solution adaptable to diverse environments.
\end{abstract}


\section{Introduction}
In recent years, the proliferation of data-driven technologies has led to significant advancements in artificial intelligence (AI) and machine learning (ML). Traditional ML models typically rely on centralized data collection, where vast amounts of data are aggregated from various sources and processed on a single server. While this approach has yielded impressive results, it also raises substantial concerns regarding data privacy, security, and ownership. With the growing importance of data privacy regulations, such as the General Data Protection Regulation (GDPR) and the California Consumer Privacy Act (CCPA), there is an increasing demand for machine learning frameworks that can protect sensitive information while still delivering robust, accurate models. Federated learning has emerged as a promising solution to these challenges, enabling the development of AI models without the need to transfer raw data to centralized servers.

Federated learning is a decentralized approach to machine learning, where the model is trained across multiple devices or servers holding local data samples, without exchanging the actual data. Instead, each participant in a federated learning system trains a model on their local data and shares only the model updates (such as gradients or weights) with a central server \cite{kairouz2021advances,li2020review,hamidi2024adafed,10381881}. This server then aggregates the updates to form a global model, which is redistributed to the participants. This process is iterated until the model converges. By keeping the data localized, federated learning addresses privacy concerns and reduces the risk of data breaches, while also enabling the model to learn from a diverse set of data sources that reflect different environments, user behaviors, or regional variations \cite{10487854}.

Beyond privacy, federated learning offers additional benefits, such as reducing the computational load on centralized servers and mitigating the need for large-scale data transfers, which can be costly and time-consuming. This decentralized approach also enables continuous learning from distributed data sources, making it highly applicable in dynamic environments where data is generated in real-time, such as mobile devices, IoT networks, or smart agriculture systems \cite{9833972}. As industries increasingly seek to harness the power of machine learning while adhering to stringent data protection standards, federated learning is poised to play a critical role in the next generation of AI applications, offering a scalable, privacy-preserving solution for a wide range of domains.

Our project is an Automated Irrigation System designed to reduce water wastage in both residential gardens and large-scale agricultural fields. Leveraging federated learning, our system aims to optimize irrigation practices by allowing local data-driven decisions while contributing to a global model that continuously improves water management strategies.

World Bank estimates suggest that 70\% of all water used globally is dedicated to agriculture, with up to 60\% potentially wasted. Optimizing crop yield while ensuring efficient resource usage requires maintaining the proper conditions for plant growth. As the global population grows, expanding agricultural capacity while minimizing water consumption has become a critical challenge. Current irrigation systems typically focus on monitoring soil moisture and water usage, but they often rely on centralized data processing, which may not account for regional environmental variations. Our system addresses this by employing a federated learning approach, enabling decentralized moisture measurement while maintaining cost-effectiveness for small-scale farmers. As climate change exacerbates the scarcity of water resources, effective irrigation management technologies are essential to sustain agricultural productivity and resource efficiency.

The primary objective of our system is to manage irrigation needs efficiently, reducing water and electricity wastage through the use of digital technologies and wireless sensors. With water becoming an increasingly scarce resource, agriculturists face growing concerns about the availability of water for irrigation now and in the future. Effective water regulation is crucial, as both insufficient and excessive watering can negatively impact farming. By implementing federated learning, our system allows local adjustments based on specific crop requirements and environmental conditions, contributing to a globally trained model that enhances long-term agricultural sustainability.

The project aims to reduce water wastage without compromising crop yield or productivity, making it adaptable for both small-scale home gardens and large-scale farms. The system comprises a soil moisture sensor for detecting soil moisture levels, an Arduino Uno microcontroller, and a motor and water pump for irrigation. By using federated learning, users are notified of moisture imbalances through a decentralized approach that respects privacy and leverages collective intelligence, ensuring optimal water usage across diverse environments.

\begin{figure}[!t]
\centerline{\includegraphics[width=8cm]{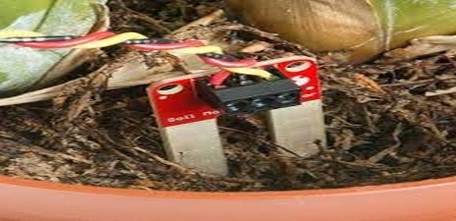}}
\caption{Soil sensor in operation}
\label{fig}
\end{figure}

\section{Literature Review}

There are various ways of detecting soil moisture by using different types of sensors available. There are two primary methods to measure the soil moisture levels - gravimetric and volumetric measurement, as described by Aniley et. al who propose the use of nanomaterials in soil sensor design for more precise detection \cite{b2}.

Many existing systems also employ different sensors for different soil quality factors and directly report the results in the form of moisture, water level,temperature etc. \cite{b3}. Some are also directly integrated with an irrigation mechanism to directly act on the input provided by the sensors. An example of this is the connection of drip irrigation mechanisms to sensor data via microcontroller for controlled irrigation \cite{b4}. An evaluation of such a system that uses controlled drip irrigation on wheat crops show higher photosynthesis rates in plants, which may be indicative of potential for a better crop yield using automation \cite{b5}. A similar system was proposed for implementation at a larger scale by Yin et. al in a Myanmar University for management of the University parks. \cite{b6}. 

Prosojo et al. improved on this system by introducing a solenoid valve for direct control of water mains for irrigation based on data sensed by the soil moisture sensor \cite{b7}.  Alternately, Mediawan proposes a system for watering house plants that provide mechanism for automation of watering as well as detecting lack of hydroponic nutrients. \cite{b8}

\section{Proposed Methodology}

Fig. 2. shows a rough outline of our proposed model. The system is a completely automatic irrigation system based on Arduino Uno microcontroller that can be divided into two major components:

\begin{itemize}
    \item Monitoring System
    \item Alerting System
\end{itemize}

\subsection{Monitoring System}

 There are two output pins on the soil moisture module employed corresponding to Digital and Analog output. Using a comparator, the output from the moisture sensor's probe is compared to a reference value. The reference value can be modified by adjusting the module's potentiometer \cite{hamidi2019systems}. When the soil is wet, the digital pin displays an active low output. The analog output from the module is used by feeding it to one of Arduino's analog pins. The flowchart in Fig. 3. describes the working of the model.

\begin{figure}[!t]
\centerline{\includegraphics[width=9cm]{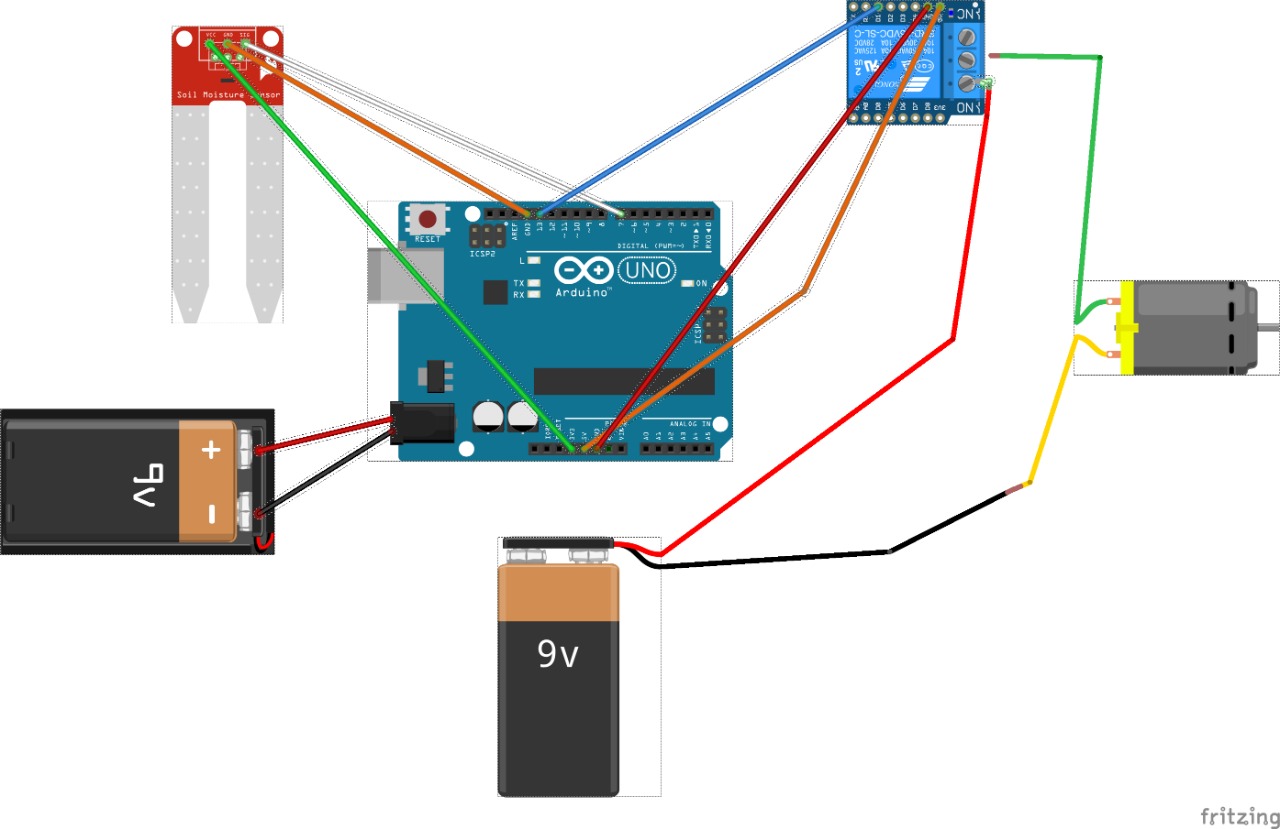}}
\caption{Proposed System}
\label{fig}
\end{figure}

The wet detection value can be modified in the program while using the analog output. A switch is connected to one of the analog pins of Arduino and a resistor is used to pull up the line. Arduino's analog pins can be used as digital inputs. The output of the switch can be used to determine the status of the tank. For measuring the level of water in the tank, Arduino reads the voltage dropped across the resistor.

\begin{figure}[!t]
\centerline{\includegraphics[height= 15cm, width=7cm]{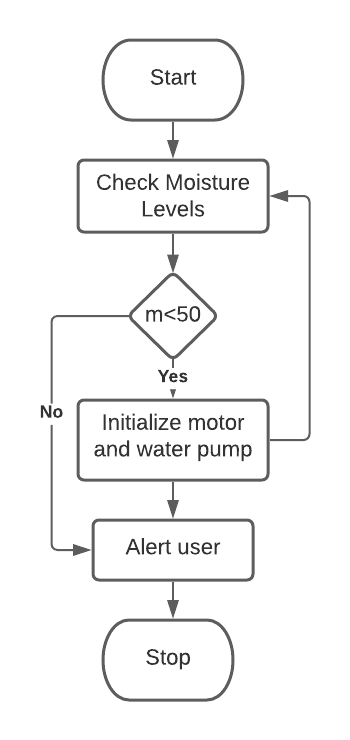}}
\caption{Flowchart of the proposed model}
\label{fig}
\end{figure}

\subsection{Alerting System}

Based on the detected values from the sensor, the user is notified of the soil moisture levels. This data can be used by the user to constantly monitor the soil conditions and assess the external factors that contribute to varying soil moisture levels while also appropriately managing the moisture content without overusing water resources. A precise account of exact water requirements of soil ensures that the user does not over or under water the plants.

A relay and motor are linked to the Arduino. A small water pump is controlled by the relay. A transistor connected to the digital pin 11 of Arduino is used to drive the system. For the system to pump water, the motor requires about 5v to 12v.

To notify the status of the soil moisture, we have used the AWS mechanism. These are the steps we followed to notify for the status of the soil moisture:
\begin{itemize}
  \item Installation of boto3 library of the AWS toolkit was done in a python supported IDE. We used Jupyter.
  \item Created an account with AWS and verified the account.
  \item Logged in to the verified AWS account and searched for the Identity and Access Management service (IAM).
  \item Added a user and activated the AmazonSNSFullAccess permission.
  \item The added user receives the user id and password.
  \item AWS is configured using the terminal which asks for the user id and password provided and also the default region name is configured.
  \item Next in the AWS services the Simple Notification Service is activated where the text messaging preferences are set and the desired phone numbers are added and then verified in the "Sandbox destination phone numbers".
  \item Lastly a python code is written using the boto3 library to check if the the moisture level is between what range and the corresponding status is sent to the verified phone numbers.
\end{itemize}

\section{System Design}
The system uses an Arduino Uno integrated with a soil sensor, motor and water pump. The sensed data is then used to alert the user using AWS messaging services.

The system consists of the following components:

\begin{itemize}
    \item Arduino UNO
    \item Soil Moisture Sensor
    \item Relay Module
    \item DC Motor
\end{itemize}

\subsection{Arduino UNO}
Arduino board uses a variety of microcontrollers and microprocessors with input and output pins, both analog and digital as shown in Fig. 4. These are in turn used for interfacing with other boards or circuits. It is an open-source platform with easy to use hardware and software. The board is able to read inputs from sensors that can be easily integrated with the board, such as temperature sensor, moisture sensor etc. It then processes and actuates the output, such as by displaying it on an LCD screen or activating a motor.

\begin{figure}[!t]
\centerline{\includegraphics[width=8cm]{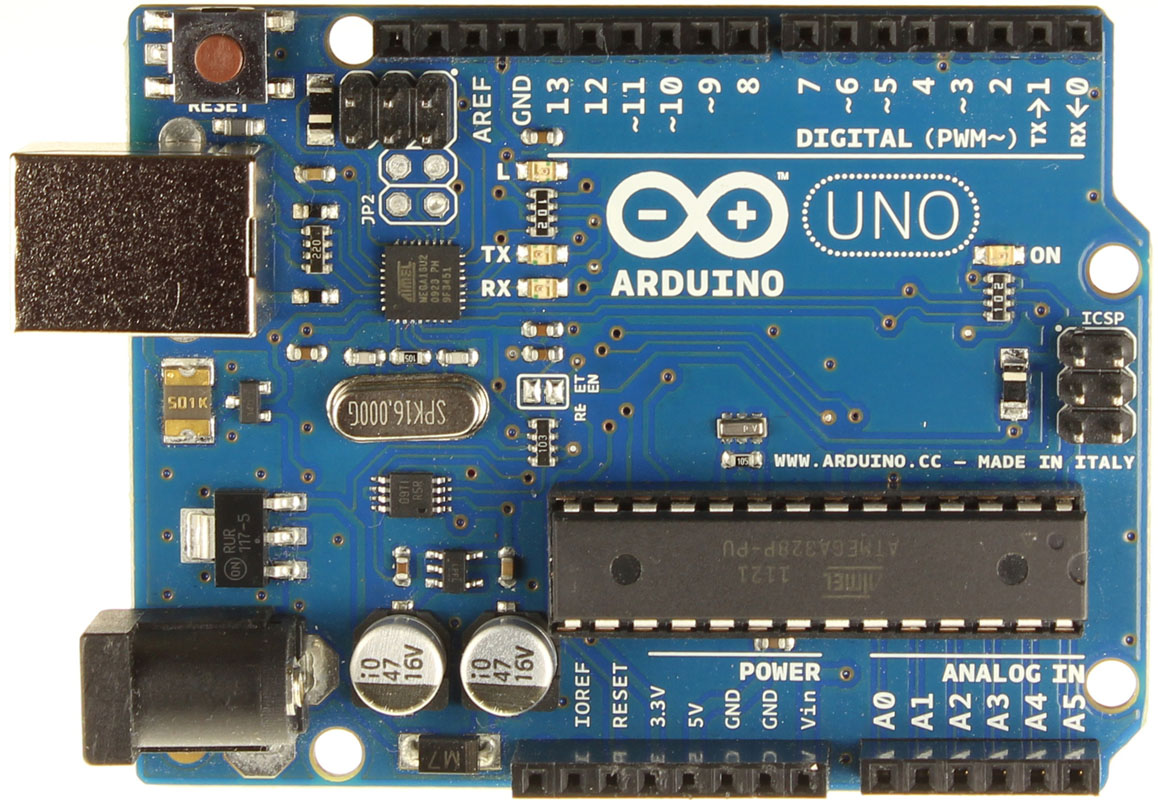}}
\caption{Arduino UNO}
\label{fig}
\end{figure}

\subsubsection{Arduino Architecture}

Arduino processor uses the Harvard architecture i.e the program code and data have separate memory. Hence, it consists of two memories: 

\begin{itemize}
    \item Program memory
    \item Data memory
\end{itemize}

The flash program memory is used to stored the program code while the data memory is used to store the data. The flash memory of Atmega328 has 32KB memory for storage of code, of which the bootloader uses 0.5KB. SRAM and EEPROM use 2KB and 1 KB respectively. The clock speed of the device is 16MHz

\subsubsection{Arduino Pins}

Fig. 5. describes the Arduino pin diagram and architecture. Arduino Uno consists of an ATmega328, a 28 pin microcontroller. These pins include 6 analog input/output pins, 14 digital input/output pins. Clock speed of 16 MHz is maintained by a crystal oscillator. The board also includes a USB connection, power jack, ICSP header and a rest button.

\begin{figure}[!t]
\centerline{\includegraphics[width=8cm]{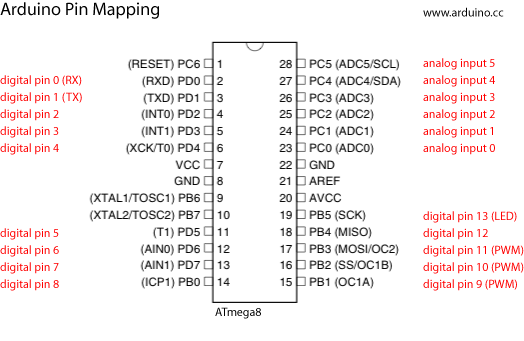}}
\caption{Arduino Pin Diagram}
\label{fig}
\end{figure}

\begin{enumerate}
    
    \item Digital Pins:
    Arduino Uno contains 14 digital input/output pins which operate on 40mA current. Many of this pins also have special functions. 6 of these pins i.e. pins 3, 5, 6, 9, 11 are PWM pins. They can be used to produce Pulse Width Modulation outputs. Pins 0 and 1 act as Receiver and Transmitter for serial communication. Pins 2 and 3 can act as external interrupts and Pin 13 is connected to an LED.
    
    \item Analog Pins:
    It has 6 analog input/output pins of resolution 10 bits.
    
    \item Power Jack:
    Arduino can either directly be powered by the PC through a USB or via an external source such as adaptors or batteries. It operates on an external supply of 7-12V. Power is applied externally to the $V_{in}$ or by giving voltage reference via the IO Ref pin.
    
    \item ARef:
    It provides reference to the analog inputs
    
    \item Reset:
    Used to reset the microcontroller.
\end{enumerate}

\subsection{Soil Moisture Sensor}

The dual probe soil moisture sensor is used to measure the moisture content of the soil. The two probes allow an electric current to travel through the soil and measure the moisture content of the soil. The soil moisture sensor is connected to an amplifier as described in Fig. 6.

\begin{figure}[!t]
\centerline{\includegraphics[width=8cm]{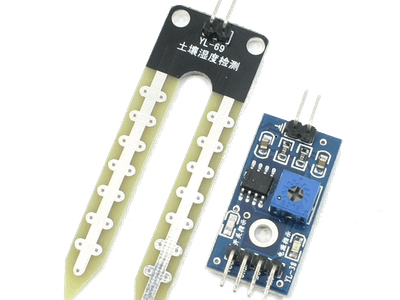}}
\caption{Soil Moisture Sensor}
\label{fig}
\end{figure}

The calculation of moisture levels is based on the resistance offered by the soil to the current. This is used the property of conductivity of electricity for water. When the moisture content in the soil is high, the conductivity of the current is increased and hence, it conducts more electricity. Similarly, lack of moisture in soil increases its resistivity  which leads to less conduction of electricity  \cite{b9}. Hence, current conducted between the probes is used to identify the soil moisture content.

When there is more water, the soil conducts more electricity, which means that the resistance will be less. So the moisture level will be higher. Dry soil reduces conductivity. So, when there is less water, the soil conducts less electricity, which means it has more resistance. So the moisture level will be lower.

\subsubsection{Soil Moisture Sensor Pins}

The soil moisture sensor pin diagram is given in Fig. 7. Soil moisture sensor consists of the following pins:

\begin{figure}[!t]
\centerline{\includegraphics[width=10cm]{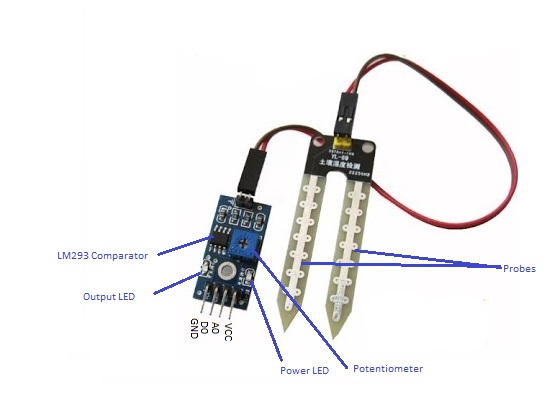}}
\caption{Soil Moisture Sensor Pins}
\label{fig}
\end{figure}

\begin{enumerate}
    \item $V_{CC}$
    \item GND
    \item Analog Out: The output at the AO pin varies with the moisture content  the soil.
    \item Digital Out: The DO pin produces a binary output based on a threshold level of moisture. If the moisture content of the soil is less that the threshold, the output at DO is '1', else it is '0'. The value of threshold can be adjusted using a potentiometer. A comparator is used for measuring the moisture content against the threshold level.
\end{enumerate}.

\subsubsection{Sensor Installation}
The sensor is directly buried in soil for moisture detection where the following should be kept in mind:

\begin{itemize}
    \item The soil surrounding the sensor should represent the entire irrigation land.
    \item Sensors need to be buried near the plant roots to be irrigated. Since plants extract water from the root zone, moisture estimation in this region ensures better water management. 
    \item The sensor should typically be buried three inches below soil surface level.
    \item Sensors must be firmly packed in the soil without excessive or less soil. There should not be any air gaps around the probes.
    \item If a single sensor is being used to irrigate a large area, it must be placed in the zone that has maximum water needs such as the area with maximum sun exposure.
    \item Sensors must be buried at least 5 feet from a home, property line, or any impervious surface.
    \item They must also be located about 5 feet from the irrigation system and closer to the center of the irrigation land.
\end{itemize}

\subsection{Relay Module}
A relay is an electrically actuated switch that is turned on or off, to allow the flow of current. It can be controlled using low voltages such as the 5V provided by the Arduino pins \cite{b10}. The module used in this project consists of 2 channels as shown in Fig. 8. Relay module contains the following pins:

\begin{figure}[!t]
\centerline{\includegraphics[width=7cm]{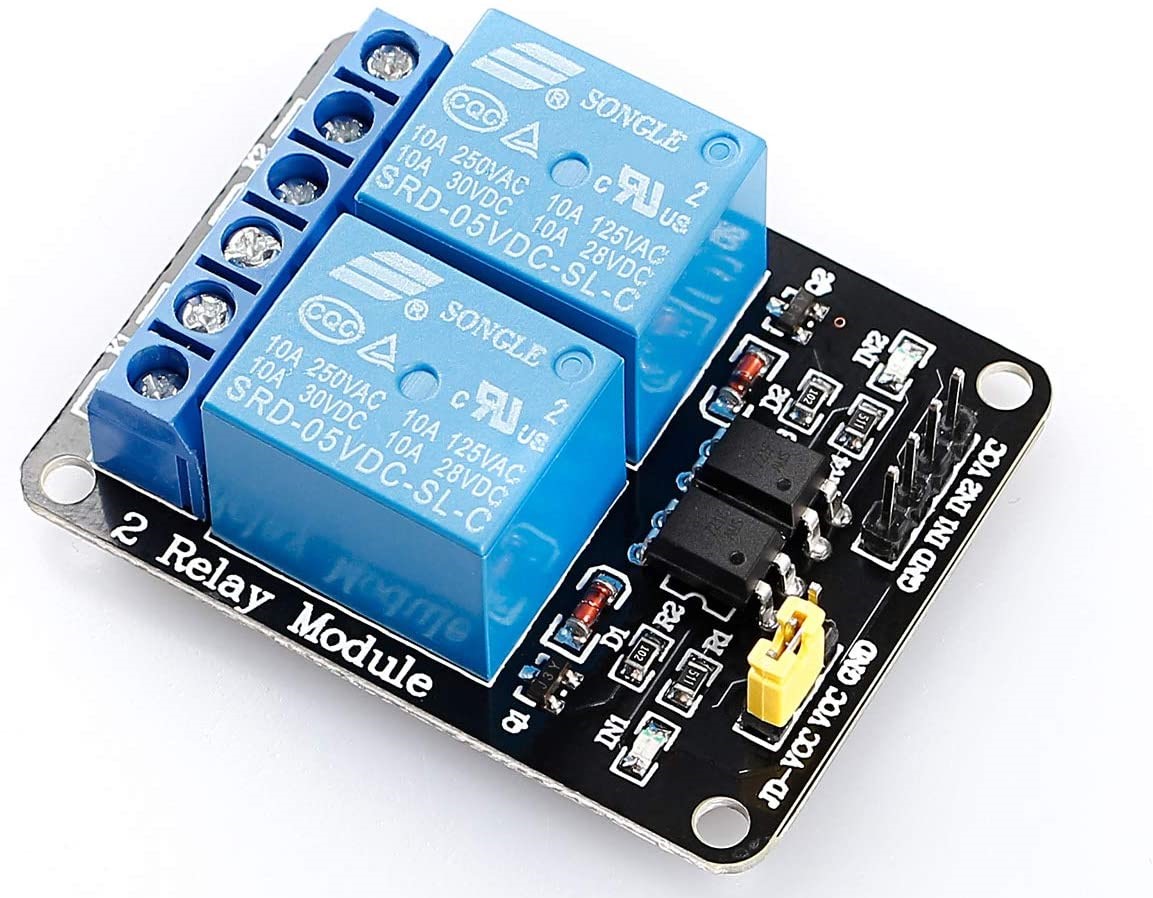}}
\caption{Relay Module}
\label{fig}
\end{figure}

\begin{enumerate}
    \item $V_{CC}$
    \item GND
    \item COM: Common pin
    \item NO (Normally Open): The normally open pin is usually not connected to the common pin. When the relay is triggered, the NO pin connects to the COM pin to provide supply to load.
    \item NC (Normally Closed): The normally closed pin is usually connected to the common pin. When the relay is triggered, the circuit is opened and no supply is provided to the load.
    \item IN1: Controls First relay. Connected to Arduino digital pin.
    \item IN2: Controls Second relay
\end{enumerate}

\subsection{DC Motor}

A DC motor is used to irrigate the field from a water source based on the sensed data. It is built around an IC1 555 Timer as shown in Fig. 10. The motor is also used to control the water level in the tank by setting a threshold low and maximum capacity.

\begin{figure}[!t]
\centerline{\includegraphics[width=7cm]{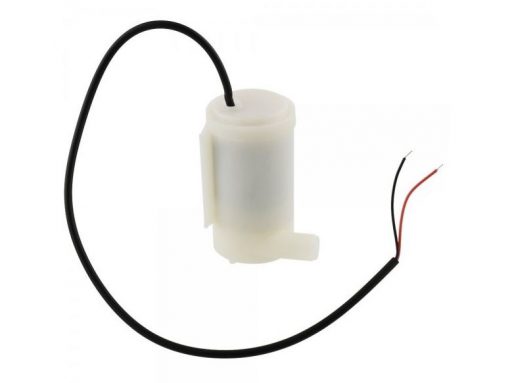}}
\caption{DC Motor}
\label{fig}
\end{figure}

\subsubsection{Water level below threshold}
If the level of the tank goes below the threshold, the voltage at IC1 pin 2 becomes low. This result in the SR flip-flop of IC1 making its output high at pin 3. The output activates the relay driver transistor T1 (BC547) and energises relay RL1. Power supply for the water resource storage provided through n/o contacts of the relay is powered on. This results in water filling in the tank

\subsubsection{Water level above maximum}
Once the level of the tank reaches the maximum capacity level, the voltage at IC1 pin 6 becomes high. This results in the SR flip-flop of IC1 making its ouput low at pin 3. Hence, this output cuts off relay driver transistor T1 and de-energises RL1. The power supply for the water tank is disconnected through the n/c contacts and of the relay. The tank stops being filled with more water

\begin{figure}[!t]
\centerline{\includegraphics[width=8cm]{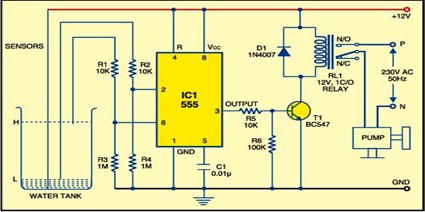}}
\caption{Working of DC Motor}
\label{fig}
\end{figure}

\section{Implementation}
The system operates in the following order:

\begin{enumerate}
    \item Programming the Arduino
    \item Sensing the soil moisture
    \item Switching the motor on and off
    \item Notifying the user
\end{enumerate}

\subsection{Programming the Arduino}

First we connect the Arduino UNO to a PC that has the Arduino IDE software installed in it. The code is executed on the Arduino IDE and sent through USB to the Arduino UNO. The Arduino microcontroller is responsible for the entire irrigation automation. The soil sensor output is fed into the D7 digital pin of the Arduino \cite{b11}. We then program the Arduino board to extract the soil moisture sensor data to a csv file \cite{b12}.

\begin{figure}[!t]
\centerline{\includegraphics[width=9.5cm]{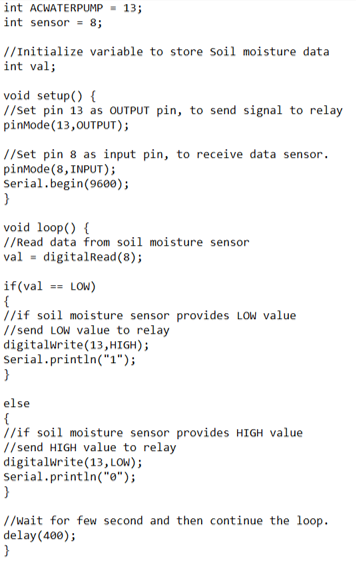}}
\caption{Code to recover data from Arduino}
\label{fig}
\end{figure}

\begin{verbatim}
  BEGIN
   Initialize variable to store moisture data
   Initialize sensor to pin 8
   Set pin 8 as input to receive data
   
   Initialize DC motor to pin 13
   Set pin 13 as output to send data to relay

   Read data from soil moisture sensor

   If soil moisture is low:
      then send low value to relay
   Else
      send high value to relay
   Repeat
END 
\end{verbatim}

The soil moisture sensor consists of two primary components:

\begin{enumerate}
    \item Two probe lead sensor
    \item Amplifier A/D circuit
\end{enumerate}

The probes go into the soil while the amplifier circuit is connected to the Arduino. The amplifier circuit contains the following pins:

\begin{itemize}
    \item $V_{in}$
    \item GND
    \item Analog data pins
    \item Digital data pins
\end{itemize}

The presence of both analog and digital outs ensures that we can recover the sensed data in either form. The sensor is designed such that it recovers the soil moisture content based on the dielectric constant of the soil \cite{b13} The dielectric constant is the ability of the soil to conduct electricity and hence, is directly proportional to the moisture content of the soil. This is because the dielectric constant of water is much greater than soil components and hence addition of water to the soil greatly increases its conductivity. Hence, it is a sufficient measure to estimate the soil moisture levels. In order to integrate the sensor with the Arduino \cite{b14,struhsaker2020methods}:

\begin{enumerate}
    \item Connect the two pins from the Sensor to the two pins on the Amplifier circuit via hook up wires.
    \item Connect the $V_{cc}$ from the Amplifier to the 3.3V pin of the Arduino
    \item Connect the GND pin to the GND pin of the Arduino.
    \item Connect the Digital data pin to the D7 pin of the Arduino to collect Digital data.
\end{enumerate}

\begin{figure}[!t]
\centerline{\includegraphics[width=9cm]{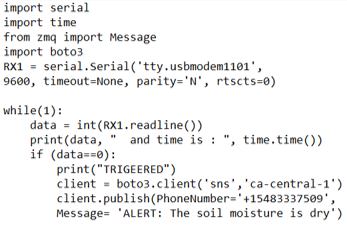}}
\caption{Code for notifying user}
\label{fig}
\end{figure}

\subsection{Switching the motor on and off}
A relay driven by a transistor connected to D11 digital pin or Arduino is used to activate a small motor. It requires a power supply of 5-12V to drive the motor. The water pump also senses and regulates the water level of the tank. This is done using an IC1 555 timer by setting a threshold and maximum water level. Based on these levels, the power supply to the water source for the tank is moderated \cite{b15}.

\subsection{Notifying the user}

\begin{figure}[!t]
\centerline{\includegraphics[width=8cm]{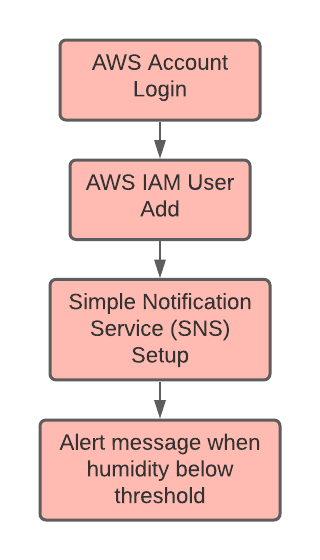}}
\caption{Flowchart for user notification}
\label{fig}
\end{figure}

\begin{enumerate}
    \item The notification system is configured using AWS. 
    \item Firstly, a python suitable IDE is downloaded. We have downloaded in the Anaconda Navigator.
    \item Next, boto3 library from the AWS toolkit is installed in the IDE.
    \item An AWS account is created to access the AWS services.
    \item The services like Identity and Access Management (IAM) and Simple Notification Service (SNS) is used to implement the text messaging services.
    \item Added a user and activated the AmazonSNSFullAccess permission.
    \item The added user receives the user id and password.
    \item AWS is configured using the terminal which asks for the user id and password provided and also the default region name is configured.
    \item Next in the AWS services the Simple Notification Service is activated where the text messaging preferences are set and the desired phone numbers are added and then verified in the "Sandbox destination phone numbers".
\end{enumerate}
    
\begin{figure}[!t]
\centerline{\includegraphics[width=9.5cm]{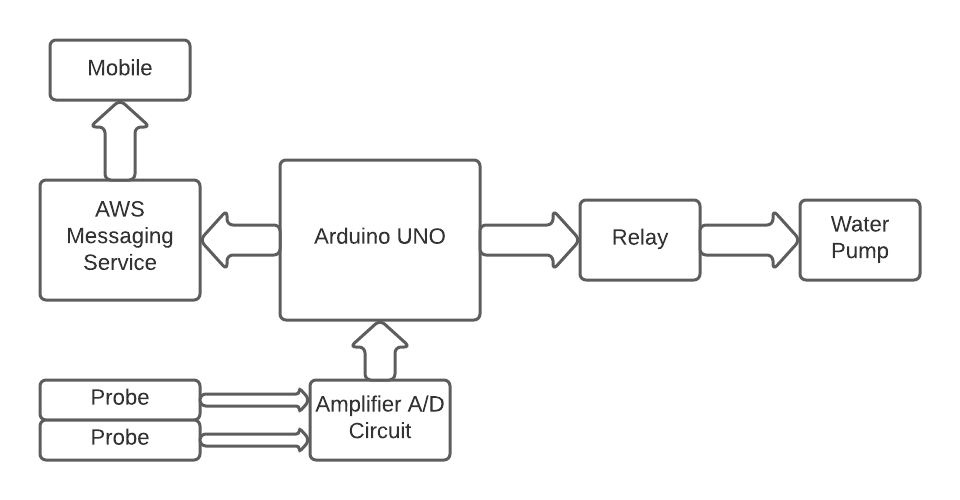}}
\caption{Block Diagram}
\label{fig}
\end{figure}

\section{Result}

We have integrated all the components into a system to fully automate the sensing and irrigation process. The sensor probes feed binary moisture data to the Amplifier circuit which is subsequently given to the Arduino. Based on this result, the water pump is activated to water the plants and the user is notified of the soil condition.

\begin{figure}[!t]
\centerline{\includegraphics[width=8cm]{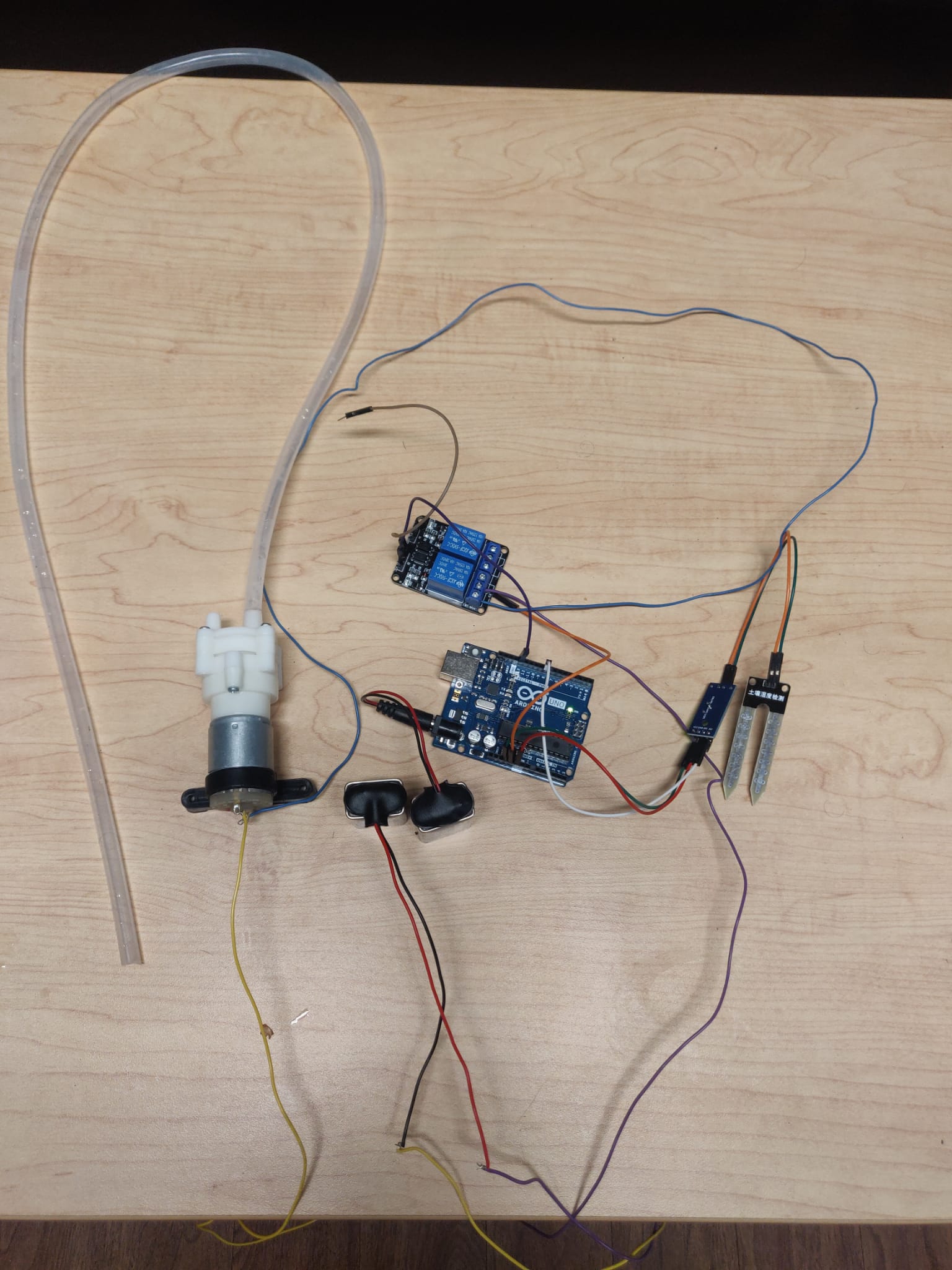}}
\caption{System Apparatus}
\label{fig}
\end{figure}

As seen in Fig. 16, the user is notified by a message stating ALERT: The soil moisture is dry to inform them that the soil requires watering. By tracking these alerts, the user can form a better understanding about the external factors that are negatively affecting the soil moisture content and consequently make more informed decisions on how to improve soil quality.

\begin{figure}[!t]
\centerline{\includegraphics[width=9cm]{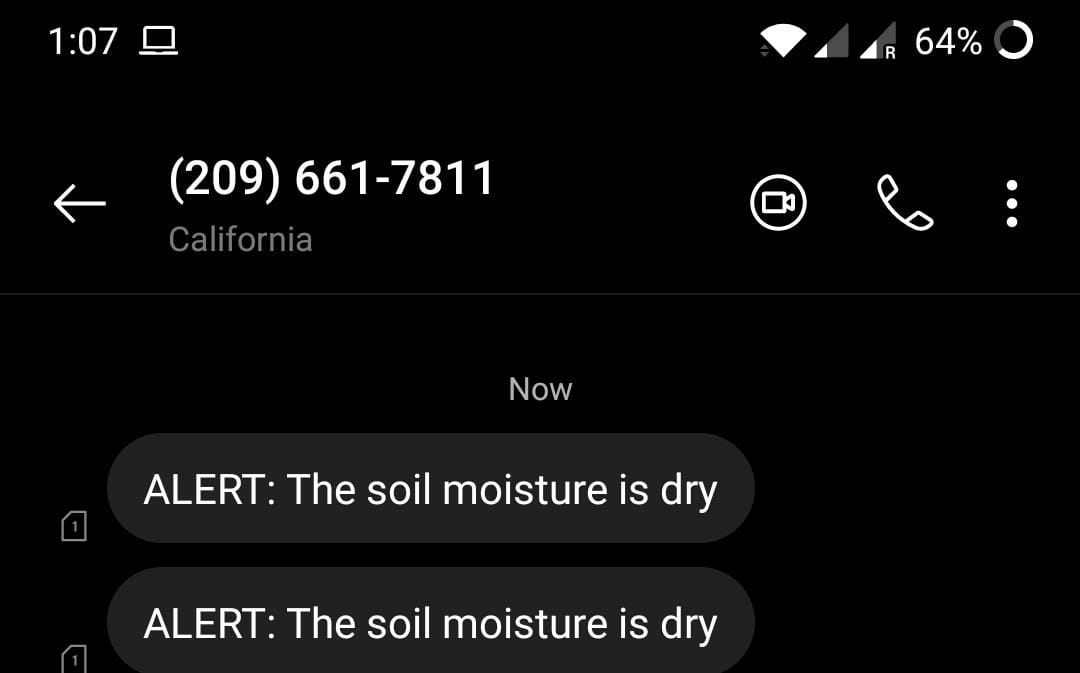}}
\caption{Alert given to user}
\label{fig}
\end{figure}

\section{Advantages of Agricultural Automation}
\begin{enumerate}
    \item Contributes greatly to water conservation by only watering plants when there is a lack of moisture in the soil.
    \item The framework ensures minimal requirement for human intervention after installation. Hence, once installed the system can more or less maintain the quality of field water levels without much effort.
    \item Due to the simplicity of the system, it does not require much expertise to install and operate.
    \item Given the low cost of components, the system is economical and hence has large scale applications in both home gardens and large scale agricultural farms.
    \item Studies show that the use of such automated systems can highly increase the crop yield and quality due to constant monitoring, care and maintenance
\end{enumerate}

\section{Future Work}
The system can be improved by adding a rain gun sensor, such that we can prevent flooding when it rains. The model can also be modified to analyze groundwater resources for better crop management. The system can also be modified by the addition of additional sensors such as temperature and water level sensors for more close real time monitoring of plant conditions. The low cost of components and ease of management mean that any improvements to the system by integrating multiple sensors are economical and effective in managing the needs of the farm.

\section{Conclusion}

Constant monitoring of soil moisture status and a sufficient knowledge of the status of the soil is crucial for effective management of crops with limited water resources. Hence, automation of irrigation system for agricultural purposes can be an economical and profitable endeavor. In terms of power consumption and hardware components used, the system is very efficient. The system is suitable for small scale home gardens to ensure operation without human intervention but its applications can also be extended to large scale farms to precisely manage water levels in order to increase crop quality and yield. This is especially useful in the case of non-uniform water distributions to spatially manage the field by differentiating irrigation methods for different soil types. Not only is this advantageous to the agriculturists, it is also very useful for water conservation efforts and hence contributes positively to the environment. In case of the use of saline water, its applications can also be extended to monitor salinity of the soil and its effects on the crop yield.

\bibliographystyle{plain}
\bibliography{mybib}

\end{document}